\documentstyle[11pt]{article}
\newcommand{\be}{\begin{eqnarray}}
\newcommand{\ee}{\end{eqnarray}}
\title{\begin{flushright}
{\small SUNY-NTG-94-11}
\end{flushright}
{\bf THE INSTANTON DENSITY AT FINITE TEMPERATURES }}
\author{
 {\bf E.Shuryak and M.Velkovsky} \\
{\it Physics Department,
State University of New York, Stony Brook, NY 11794}}

\begin{document}

\maketitle
\centerline{\bf Abstract}
  For {\it low} T  new strict results for the instanton density n(T)
 are reported. Using
the PCAC methods, we express n(T)
in terms of {\it vacuum}
average values of certain operators, times their {\it calculated} T-dependence.
At high T, we discuss the {\it applicability} limits
of the perturbative results. We further speculate about possible behaviour
of n(T) at $T\sim T_c$.

\newpage
\section{ Introduction}
  Tunneling between topologically different configurations of the gauge
field, described semiclassically by instantons \cite{BPST},  dominate
 the physics of light quarks. In early works (summarized e.g. in
\cite{CDG}) instantons were treated as a dilute gas, while later it was
recognised that the instanton
ensemble resemble rather a strongly interacting ``instanton liquid"
 \cite{Shuryak_82,DP}.
 During the last years calculations of
the correlation functions \cite{SV} and Bethe-Salpeter wave functions \cite{SS}
for various mesonic and baryonic channels were made
along these lines. The results agree surprisingly well both with
 phenomenology
 \cite{Shuryak_cor}
and lattice simulations \cite{Negele}. Parameters of the ''instanton liquid"
 were also reproduced
(by  the ''cooling" method) directly from the (quenched) lattice configurations
 \cite{Negele_DALLAS}.
 In addition to that, it was found that correlation functions
as well as hadronic wave functions in most channels remain
practically unchanged after ''cooling". In particular, main mesons and e.g.
the nucleon remains bound, with about the same mass and wave function.
This confirms that the
agreement of previous lattice calculations with the instanton model
was not accidental, and instantons indeed are
the most important non-perturbative phenomena in QCD.

   Investigations of  the finite temperature case were started in \cite{HS},
where classical {\it caloron}
 solution was found. Although the solution depends on T,
the action is T-independent. Furthermore, it was argued
by one of us \cite{Shuryak_conf}
that at high T the specific charge renormalization and the
 Debye-type screening of the electric field in quark-gluon
plasma should suppress instantons with size $\rho > 1/T$.  Pisarski and Yaffe
\cite{PY} have evaluated the T-dependence in the one-loop approximation.
The physical nature of their result and its applicability region will be
discussed below.

   Last years the studies of the finite temperature
had focused especially on the region around  the chiral
phase transition $T\approx T_c$. The first attempt to understand this
phase transition as a rearrangement of the instanton liquid, going from
a random phase at low temperatures to a strongly correlated ``molecular"
phase at high temperatures was made in \cite{IS}. Recently
 this idea was recently made more quantitative in \cite{IS2,ssv},
and although the detailed comparison to lattice thermodynamics and correlation
functions is yet to be made, the first results show overall agreement,
indicating that the mechanism of chiral restoration is basically
understood.

   The particular topic of this paper, the {\it temperature
dependence of the instanton
density} n(T), is certainly an important ingredient of all this development.
However, it has attracted surprisingly little attention in literature.
The only attempts to determine this quantity by ``cooling" of the lattice
configurations (same method  as was used at T=0) were made in
refs.\cite{lattice}. The only statement one can probably
make using these data is that the density has
{\it no} significant T-dependence,  till  $T\sim T_c$.
Unfortunately,
 accuracy  of this statement remains at the level of 50\%, at best.

\section{ Instantons at low temperatures}

 As far as we know, the modification of the instanton density in this limit
was never considered before.
However, the general physical picture at low $T<<T_c$ is well
known:  the heat bath is just a dilute gas of the lightest hadrons,
the pions. The problem is especially clear in the chiral limit, in which
quark and pion masses are neglected, and this case is assumed in what follows.
  Due to their Goldstone nature, the large wavelength pions are nothing else
but small
collective distortions of the quark condensate. Therefore, one can always
translate the average values  over the {\it
pion state}
as the {\it vacuum expectation value} (VEV) of a different but related
quantity.

  Let us now consider a tunneling event, described semiclassically
by the instanton solution.
As discovered by 't Hooft \cite{Hooft}, it can only take place if certain
rearrangements in the fermionic sector are made, which can be described by
some effective Lagrangian with $2 N_f$ fermionic legs.  For
simplicity, in this work we assume that  the number
of light flavors $N_f=2$, disregarding strange and heavier quarks.

  The situation can be further simplified by consideration of
  {\it small-size} instantons $\rho \Lambda_{QCD}<<1$, for which
this Lagrangian
can be considered as a {\it local} operator.
What follows from 't Hooft Lagrangian, after averaging
over the instanton orientations is made, is \cite{SVZ_79}:
\be
\Delta {\cal L}= \int d\rho d_0(\rho)({4 \over 3}\pi^2 \rho^3)^2\{\bar q
\Gamma_+ q
\bar q \Gamma_- q+{3\over 32}\bar q \Gamma_+^a q \bar q \Gamma_-^a q-
{9\over 128}\bar q
\Gamma_+^{a\mu \nu} q \bar q \Gamma_-^{a\mu \nu} q\}
\ee
where the definition of  the  operators involved is as follows
\be
\Gamma_{\pm}=({1-\gamma_5 \over 2})\otimes ({1\pm\tau_3 \over 2})
\ee
\be
\Gamma_{\pm}^a=({1-\gamma_5 \over 2})\otimes ({1\pm\tau_3 \over 2})\otimes t^a
\ee
\be
\Gamma_{\pm}^{a\mu \nu}=({1-\gamma_5 \over 2})\sigma_{\mu \nu}\otimes
({1\pm\tau_3 \over 2})\otimes t^a
\ee

At T=0 the instanton density is therefore proportional to the VEV of
$\Delta {\cal L}$, and the only thing which changes at low T is clearly the
modification of the quantities above.

 How to do this
technically was actually clarified by PCAC-related paper in 60's.
  The necessary formulae   can be found
e.g. in the recent paper by Eletsky \cite{Eletsky}, where
 different set of four-fermion
operators (appearing in QCD sum rules for vector and axial currents)
was considered.
   The general expression  is
$$
\langle (\bar q A q)(\bar q B q)\rangle _T=
\langle (\bar q A q)(\bar q B q)\rangle _0
-{ T^2 \over 96 F^2_\pi}\langle (\bar q\{\Gamma^a_5 \{\Gamma_5^a A\}\} q)
(\bar q B q)\rangle _0
$$
\be
-{ T^2 \over 96 F^2_\pi}\langle (\bar q A q)(\bar q\{\Gamma^a_5
\{\Gamma^a_5 B\}\} q)\rangle _0
-{ T^2 \over 48 F^2_\pi}\langle (\bar q \{\Gamma^a_5 A\} q)(\bar q
\{\Gamma^a_5 B\} q)\rangle _0
\ee
where A,B are arbitrary flavor-spin-color matrices and $\Gamma_5^a=\tau^a
\gamma_5$.
Here and below flavor matrices are shown as $\tau_a$, and color one
as $t^a$.

  For generality, there are six different operators
of different spin-flavor-color structure
involved (see
the Table). Their  T-dependence in $O(T^2)$ order can be found from the
expression above, and it is also listed in the Table.
Generally, the operators mix, and it is convenient to group those combinations
which do not. Returning to the effective Lagrangian, one can see
that there are only two combinations which
are actually relevant
\be
K_1={\cal O}^A_1+{3\over 32}{\cal O}^B_1-{9\over 128}{\cal O}^C_1 \\
K_2={\cal O}^A_2+{3\over 32}{\cal O}^B_2-{9\over 128}{\cal O}^C_2
\ee
The final result for the instanton density
at low temperature T  therefore contains two constants, the
{\it vacuum} averages of these operators
\be
 dn(T) = {d\rho \over \rho^5} d(\rho) ({4\over 3}\pi^2 \rho^3 )^2
\left[\langle K_1\rangle _0{1\over 4}(1-{T^2 \over 6 F^2_\pi})
-\langle K_2\rangle _0{1\over 12}(1+{T^2 \over 6 F^2_\pi})\right]
\ee
Although  two VEV's which appear here are unknown (and are subject to
further investigations), it is clear that
the total T-dependence should be {\it rather weak}: it is bound to be
$ 1 + a {T^2 \over F^2_\pi}$ with a in the
strip  $a=(- 1/6,1/6)$.

    The so called {\it vacuum dominance}  (VD)
hypothesis \cite{SVZ_79} was used
   in various applications (such as QCD sum rules and weak decays) for
evaluation of VEV's of various operators. It leads to VEV's and the
$O(T^2)$  corrections also indicated in the Table 1. Remarkably enough
\footnote{
(However, the reader should be warned that in general this approximation is
not supposed to hold or
be very accurate, particularly for the operators considered, which are related
with instantons.
},
in this case the T-dependence {\it exactly cancel}.

  Returning to discussion of our general result (8), we comment that
   this result by itself rules out some possible picture of low-T vacuum
structure. In particular, the so called ``random instanton liquid model"
(RILM) was shown to be a  reasonable approximation for the T=0 case \cite{SV}.
One may wander if the same model can describe the T-dependence  at low T.
If it is the case, the quark condensate
 should scale with the instanton density as $n_{inst}^{1/2}$, see
\cite{Shuryak_82}. Being combined with the well known chiral
theory  result
\be
 \langle \bar q q\rangle _T=\langle \bar q q\rangle _0 (1 - {T^2
\over 8 F^2_\pi})
\ee
 these two formulae lead to\footnote{
Note that the same result can be obtained by a naive assumption, which was used
in some works on QCD sum rules
in the past: namely, that average of {\it all} four-fermion operators
have T-dependence as the {\it square} of the condensates.}
$ n_{inst}(T)/n_{inst}(0)=(1 - {T^2 \over 4 F^2_\pi})$.

  However, as this estimate happen to be {\it outside}
of the
strip indicated above, this possibility is definitely {\it
ruled out}. It means that, even if RILM is a
perfect model at T=0, it cannot be so for even small T. This conclusion
agrees very well with other studies of the instanton ensemble, such as
\cite{IS2}, which emphasize the role of correlations built up with growing T
in the ensemble of instantons.

\section{ Instantons at high temperatures}

   QCD vacuum at high temperatures undergoes a phase transition into a new
phase, called the {\it quark-gluon plasma}\cite{Shuryak_QGP}. Although
 {\it virtual} gluons {\it antiscreen}
the external charge (the asymptotic freedom), the {\it real} gluons of the
perturbative heat bath {\it screen} it, leading to the well known
expression for the Debye screening mass \cite{Shuryak_QGP}:
\be
M_D^2=  (N_c/3+N_f/6) g^2 T^2
\ee
where $N_c,N_f$ are the numbers of colors and flavors, respectively.
''Normal" O(1) electric fields are therefore screened at distances $1/gT$,
while stronger ''non-perturbative" fields of the instantons $O(1/g)$
should be screened already at scale 1/T.

  Quantitative behaviour of the
instanton density at high temperatures was determined in ref. \cite{PY}.
\be
dn(\rho,T)= dn(\rho,T=0)\exp \left\{ -{1\over 3}\lambda^2(2N_c+N_f)-12
A(\lambda)\left[1+{1\over 6}(N_c-N_f) \right] \right\}
\ee
where $\lambda=\pi\rho T$, and $A(\lambda)=-{1\over 36}\lambda^2
+o(\lambda^2)$.
Therefore, at high temperatures the contribution of small size
instantons such as $T>> 1/\rho$  is
exponentially  suppressed. As a result, the
instanton-induced contribution to physical
quantities like energy density (or pressure, etc) become of the order
of\footnote{Here we consider only the pure glue theory. In the theory with
massless fermions individual instantons are impossible, and only
``instanton-antiinstanton" molecules can appear at high temperatures.}
\be
\epsilon(T) \sim \int^{1/T}_0 {d\rho \over \rho^5} (\rho \Lambda)^{(11 N_c/3)}
\sim T^4 (\Lambda/T)^{(11 N_c/3)}
\ee
which is small compared to that of ideal gas $\epsilon(T)_{ideal}\sim T^4$.

 Although the Pisarski-Yaffe formula
 contains only
the dimensionless parameter $\lambda$, its applicability is limited by {\it
two separate} conditions:
\be
\rho << 1/\Lambda, \,\,\,\,\,\,\, T >> \Lambda
\ee
The former condition ensure semiclassical treatment of the tunneling,
while the latter
is needed to justify  perturbative treatment of the heat bath.
  In this section we would like to discuss applicability conditions of these
well-known results in greater details.

Our first point is that the one-loop effective action
discussed by Pisarski and Yaffe actually
consists of {\it two parts} with very different
physical origin and interpretation. To show that in the simplest case,
consider the determinant corresponding to scalar
isospin 1/2 field\footnote{The determinants of the actual quadratic
fluctuations of the {\it quark} and {\it gluon} fields (modulo the factor
corresponding to zero fermion
modes) can be expressed via the
determinants of {\it scalar} fields with isospins 1/2 and 1 \cite{PY}.}
and rewrite them as follows:
\be
\delta =Tr_T[\log({-D^2(A(\rho,T))\over -\partial^2})]-
Tr[\log({-D^2(A(\rho))\over -\partial^2})]=\delta_1+\delta_2
\ee
\be
\delta_1 =Tr_T[\log({-D^2(A(\rho))\over -\partial^2})]-
Tr[\log({-D^2(A(\rho))\over -\partial^2})]
\ee
\be
\delta_2 =Tr_T[\log({-D^2(A(\rho,T))\over -\partial^2})]-
Tr_T[\log({-D^2(A(\rho))\over -\partial^2})]
\ee
Here $Tr_T$ is a trace over all matrix structures, plus integration over
$\cal M$ -- the strip in $R^4$ with span in the $\tau$ direction of $1/T$.
$A(\rho,T)$ is the caloron field and $A(\rho)$ is the instanton field.
Two contributions introduced in this way,
$\delta_1,\delta_2$, are the origin of two terms in the resulting formula
(11).

 As it was shown in ref.\cite{Shuryak_A4}, the first term can be expressed
via the {\it forward scattering amplitude} of heath bath constituents,
 on the  instanton field.
Therefore its physical origin is clear:  $\rho^2$ comes from the
scattering amplitude, while the temperature factor $T^2$ enters via the
standard thermal integral over the particle momenta:
\be
\delta_1=\int{ d^3p\over(2\pi)^3}{1\over 2p(\exp(p/T)-1)} TrT(p,p)
\ee

  Let us show how it works using the  example of a ``scalar quark",
which is simpler than realistic spinor and vector particles considered in
\cite{Shuryak_A4}.
One can evaluate $T(p,p)$, the forward scattering amplitude of a scalar
quark on the instanton
field, using standard  Leman-Simansik-Zimmermann reduction formula:
\be
Tr T(p,p)=\int d^4x d^4y \ e^{ip.(x-y)} Tr(\partial ^2_x \Delta_{1\over2}(x,y)
\partial^2_y)
\ee
where $\Delta_{1\over2}$ is the (isospin 1/2) scalar quark propagator
 \cite{BCCL}:
\be
\Delta_{1\over2}(x,y)={x^2 y^2+\rho^2 x.\tau y.\tau^\dagger \over
4\pi^2 (x-y)^2 x^2 y^2 (1+\rho^2/x^2)^{1\over2}(1+\rho^2/y^2)^{1\over2}}
\ee

By rescaling (18) as $\xi=px, \eta=py$, subtracting the trace of the free
propagator and going to the physical pole $p^2=0$,  one gets:
\be
Tr T(p,p)=\int d^4\xi d^4\eta e^{in.(x-y)} {\rho^2\over
2\pi^2(\xi-\eta)^2}\left({\xi .\eta \over
\xi^2 \eta^2}-{1\over 2\xi^2}-{1\over 2\eta^2}\right)=-4\pi^2\rho^2
\ee
As it is just constant, there is no problem with its analytic continuation
to small Minkowski momenta of scattered quarks, and
plugging (20) into (17) we have:
\be
\delta_1={1 \over 3}\eta\lambda^2,\ \eta= \left\{ \matrix{ 1 & for& periodic
& fields \cr -1/2 & for& antiperiodic& fields \cr}\right\}
\ee
Note also, that this scattering amplitude
has the same origin (and the same dependence
on $N_c,N_f$) as the Debye mass.

Although formally any result obtained
by  the  perturbative expansion demand smallness
of the effective charge $g(T)<<1$, it is not clear
in practice what this condition actually imply.
  However, we conjecture that accuracy of calculation sketched above
is controlled by  the same effects as the
accuracy of {\it perturbative calculation of the
Debye mass} by itself.
If so, one can use available lattice studies
of the screening phenomena (e.g.\cite{screening}) and check at which T
their results start to agree with the perturbative formula (10).
We then conclude, based on available lattice data, that
Debye mass and instanton suppression formula (11) should be valid
above
 $T> T_{pert}=3 T_c \approx 500$ MeV).

  How strong can this suppression be, at that point? Using a canonical
``instanton liquid" size of the instanton $\rho\approx1/3$ fm, one gets
suppression on the level $10^{-3}$, from the $\delta_1$ term alone. It suggests
a very dramatic behaviour in the interval from 1 to 3 $T_c$.

  Let us further speculate about
the magnitude of $\delta_1$ contribution for lower
temperatures. In the interval between $T_c$ and $T_{pert}$
it is expected on general ground (and observe on the lattice) that
the Debye mass $M_D\rightarrow 0$ at the critical temperature: screening is
gone together with the plasma.   However, another suppression mechanism should
substitute it {\it below} $T_c$,
namely the one due to  scattering of {\it hadrons} on the instanton.
This is what  we have done above for the low-T case,in which only the
 soft pions should be included.
 At this time,
we do not know how to estimate this effect including other hadrons.

  Now we turn to discussion of the second term $\delta_2$ in (14),
$A(\lambda)$ in \cite{PY}, which was
actually first obtained by Brown and Creamer
in \cite{BC}.   At small $\lambda$ it leads to the following correction
\be
\delta_2=-(1/36)\lambda^2+o(\lambda^2)
\ee
and thus it has the same sign as $\delta_1$ and parametric magnitude, just
numerically smaller coefficient.
 (In the isospin 1 case $\delta_1=(4/3)\lambda^2$ and
$\delta_2=-(4/9)\lambda^2+o(\lambda^2).$)

   The splitting of the variation of
the effective action into two physically different contributions is
a generic phenomenon.
This second term has different physical
origin, because it is connected to
a {\it quantum correction} to the colored current, times the
{\it T-dependent variation} of
the instanton field, the difference between the caloron and the instanton.

  Thus,
the finite T
effects
not only lead to appearance of a usual (perturbative) heat bath, but they
also  modify strong ($O(1/g)$) {\it classical gauge field} of the instanton.
In Matsubara formalism this is described by the
non-linear ``interference" of the
instanton field with its ``mirror images", in the (imaginary) time direction.

  Let us
conjecture, that for $T<T_c$ this suppression mechanism is actually
irrelevant, for the following reason. It is well known that in this T domain
all gluonic correlators decay strongly with distance, because all physical
``glueballs" states are very heavy.
It should make
any interaction with the ``mirror images" (at distance $\beta=1/T$)
virtually impossible. Estimating this effect as
$\exp(-M_{glueball}/T)$, where $M_{glueball}\sim
1.6 GeV$ is the mass of the
lightest glueball, one gets even at $T=T_c\sim 140 MeV$ a suppression factor
of the order of $10^{-5}$.

\section{ Summary and discussion}

   We have studied the change of the instanton density at {\it low} and {\it
high}
temperatures.

In the former case, $T<<T_c$, we have
considered the heat bath as being
made of dilute  soft pions.
Applying PCAC methods (in the chiral limit) we have derived strict
result (8) for the instanton density n(T) at low T.
 It implies very weak T dependence, which agrees with
available lattice measurements inside their (so far rather poor) accuracy.
 It also
contradicts to some naive models, for example it shows that the
 ''random instanton liquid model", presumably a good description of the
QCD vacuum, can not be true even at low T.
Fortunately, it perfectly
agrees with the current ideas about the finite-T QCD
\cite{IS2}, pointing out that quark-induced instanton-
antiinstanton correlations are building up with T, till only
instanton - antiinstanton ''molecules" remain for $T>T_c$.

   Our discussion of the
  {\it high} temperatures can be summarized as follows.
For very high $T>T_{pert}\sim 3 T_c$ the perturbative result of Pisarski and
Yaffe \cite{PY} holds.

  Furthermore,
 we have pointed out that it consists of two parts, $\delta_1,\delta_2$,
with different
underlying physics. The
first one is directly connected to
occupation densities of quarks and gluons from
the
plasma. It is the same effect as lead
e.g.
to the Debye screening mass, so one  knows from lattice data at which T
this part of the instanton suppression can be trusted. We
also claimed that it weakens toward $T_c$, but at the same
time scattering of hadrons on instantons should  appear at  $T<T_c$, and
we do not know how to take it into account (except for soft pions).

 The second term $\delta_2$
originates from the {\it T-dependent
variation of the classical field}, coupled
to a quantum correction to the colored current. We
expect this effect to become exponentially small
for  $T<T_c$, but we do not know its T-dependence in the strip 1-3 $T_c$, where
it can be very strong.

  Finally, let us repeat once more, that
understanding of  the temperature dependence of the
instanton density  is of crucial importance for understanding
of non-perturbative
phenomena at and around $T_c$. Surprisingly little efforts has been made
to clarify this question. In particularly, we call upon
lattice community to make quantitative measurements of n(T), which can
be done by well known methods.

{\bf Acknowledgements}

One of us (E.S.) acknowledge helpful discussions with A.DiGiacomo,
M.~C. Chu and V.Eletsky.
This work is supported in part by the US Department
of Energy under Grant No. DE-FG02-88ER40388 and
No. DE-FG02-93ER40768.

\section{ Table}

\moveleft 1. cm
\vbox{\offinterlineskip
\halign{\strut \vrule \ \hfil # \hfil \ & \vrule \ \hfil # \hfil \ & \vrule
\ \hfil # \hfil \ & \vrule \ \hfil # \hfil \ \vrule \cr
\noalign{\hrule}
Operator & Coeff.  & T-renormalization & Vacuum Dominance\cr
  & in ${\cal L}$ &  &Hypothesis \cr
\noalign{\hrule}
${\cal O}^A_1=$
& ${1\over 4}$ & $\langle {\cal O}^A_1\rangle _0-$
 & ${\langle \bar u u\rangle ^2_0\over 144}(132-360{T^2
\over 12 F^2_\pi})$ \cr
$\bar q {1-\gamma_5 \over 2} q \bar q {1-\gamma_5 \over 2} q$ & &
$(3\langle {\cal O}^A_1\rangle _0+\langle {\cal
O}^A_2\rangle _0){T^2\over 12 F^2_\pi}$ &  \cr
\noalign{\hrule}
${\cal O}^A_2=$ & $-{1\over12}$ &  $\langle {\cal O}^A_2\rangle _0-$ &
${\langle \bar u u\rangle ^2_0\over 144}(-36-360{T^2
\over 12 F^2_\pi})$ \cr
$\bar q {1-\gamma_5 \over 2}\tau^a q \bar q {1-\gamma_5 \over
2}\tau^a q$ & & $(3\langle {\cal O}^A_1\rangle _0+\langle {\cal
O}^A_2\rangle _0){T^2\over 12 F^2_\pi}$ &  \cr
\noalign{\hrule}
${\cal O}^B_2=$ & ${3\over128}$ & $\langle {\cal O}^B_1\rangle _0-$
 & ${\langle \bar u u\rangle ^2_0\over 144}(-64+384{T^2
\over 12 F^2_\pi})$\cr
$\bar q {1-\gamma_5 \over 2}t^i q \bar q {1-\gamma_5 \over
2}t^i q$  & & $(3\langle {\cal O}^B_1\rangle _0+\langle {\cal
O}^B_2\rangle _0){T^2\over 12 F^2_\pi}$ & \cr
\noalign{\hrule}
${\cal O}^B_2=$ & $-{1\over 128}$ & $ \langle {\cal O}^B_2\rangle _0-$
 & ${\langle \bar u u\rangle ^2_0\over 144}(192+384{T^2
\over 12 F^2_\pi})$ \cr
$\bar q {1-\gamma_5 \over 2}t^i \tau^a q \bar q {1-\gamma_5
\over 2}t^i \tau^a q$ & & $(3\langle {\cal O}^B_1\rangle _0+\langle {\cal
O}^B_2\rangle _0){T^2\over 12 F^2_\pi}$ & \cr
\noalign{\hrule}
${\cal O}^C_1=$
& $-{9\over 512}$ & $ \langle {\cal O}^C_1\rangle _0-$ &
${\langle \bar u u\rangle ^2_0\over 144}(768-4608{T^2
\over 12 F^2_\pi})$ \cr
$\bar q {1-\gamma_5 \over 2}\sigma_{\mu \nu}t^i
 q \bar q {1-\gamma_5 \over 2}\sigma_{\mu \nu}t^i q$ &
&$(3\langle {\cal O}^C_1\rangle _0+\langle {\cal
O}^C_2\rangle _0){T^2\over 12 F^2_\pi}$ &  \cr
\noalign{\hrule}
${\cal O}^C_2=$& ${9\over 1536}$ & $\langle {\cal O}^C_2\rangle _0-$
 & ${\langle \bar u u\rangle ^2_0\over 144}(2304-4608{T^2
\over 12 F^2_\pi})$ \cr
$\bar q {1-\gamma_5 \over 2}\sigma_{\mu \nu}t^i \tau^a
 q \bar q {1-\gamma_5 \over 2}\sigma_{\mu \nu}t^i \tau^a q$  & &
$(3\langle {\cal O}^C_1\rangle _0+\langle {\cal
O}^C_2\rangle _0){T^2\over 12 F^2_\pi}$ & \cr
\noalign{\hrule}
}}

\end{document}